\documentclass[10pt,twocolumn]{IEEEtran}
\usepackage[dvips]{graphicx}

\usepackage{epsf,epsfig,array}
\usepackage{amsmath,amssymb}
\usepackage{cite}
\usepackage{picins}
\usepackage{graphicx}
\usepackage{epstopdf}
\usepackage{subfigure}
\usepackage{epsf,epsfig}
\usepackage[]{latexsym}
\makeatother
\makeatother
\renewcommand{\normalsize}

\renewcommand{\baselinestretch}{1}

\begin{document}
\title{Opportunistic Spectrum Sharing in Dynamic Access Networks: Deployment Challenges, Optimizations, Solutions, and Open Issues}
\author{Haythem Bany Salameh\\
haythem@email.arizona.edu}
\maketitle
\begin{abstract}
In this paper, we investigate the issue of spectrum assignment in CRNs and examine various opportunistic spectrum access approaches proposed in the literature. We provide insight into the efficiency of such approaches and their ability to attain their design objectives. We discuss the factors that impact the selection of the appropriate operating channel(s), including the important interaction between the cognitive link-quality conditions and the time-varying nature of PRNs. Protocols that consider such interaction are described. We argue that using best quality channels does not achieve the maximum possible throughput in CRNs (does not provide the best spectrum utilization). The impact of guard bands on the design of opportunistic spectrum access protocols is also investigated. Various complementary techniques and optimization methods are underlined and discussed, including the utilization of variable-width spectrum assignment, resource virtualization, full-duplex capability, cross-layer design, beamforming and MIMO technology, cooperative communication, network coding, discontinuous-OFDM technology, and software defined radios. Finally, we highlight several directions for future research in this field.
\end{abstract}

\section{Introduction}
The design of spectrum sharing and access mechanisms for cognitive radio networks (CRNs) has attracted much attention in the last few years. The interest in CRNs is mainly attributed to their ability to enable efficient spectrum utilization and provide wireless networking solutions in scenarios where the un-licensed spectrum bands are heavily utilized. Furthermore, due to their cognitive nature, CRNs are more spectrum efficient and robust than their non-cognitive counterparts against spectrum unavailability, and have the capability to utilize different frequency bands and adapt their operating parameters based on the surrounding radio frequency (RF) environment. Specifically, CR is considered as the key technology to effectively address the inefficient spectrum utilization in legacy licensed wireless communication systems by providing opportunistic on-demand access \cite{[1],[13]}. CR technology enables unlicensed users to opportunistically utilize the idle PR channels
(so-called spectrum holes). The spectrum holes represent the PR channels that are currently under-utilized. In order to utilize these spectrum opportunities without interfering with the PRNs, CR users should perform accurate spectrum sensing, through which idle channel
lists are identified. In addition, the CR users should be flexible enough to quickly vacate the operating channel when a PR user reclaims it. In this case, CR users should quickly and seamlessly switch their operating channel(s).

While large-scale deployment of CRNs is still to come, extensive research attempts are currently underway to improve the effectiveness of spectrum sharing protocols and improve the spectrum management and operation of such networks~\cite{[13],[16a],[3av],[5av],[116],[120],[124],[37v],[125],[37]}. Two of the most crucial challenges in deploying CRNs are the needs to maximize spectrum efficiency and minimize the caused interference to PRNs. On other words, providing efficient communication and spectrum access protocols that provide high throughput while protecting the performance of licensed PRNs is the crucial design challenge in CRNs.

The main objective of this paper is to overview and analyze the key schemes and protocols for spectrum access/sharing/managemnent that have been developed
for CRNs in the literature. Furthermore, we briefly highlight a number of opportunistic spectrum sharing and management schemes and explain their operation details. As indicated later, it follows logically that cross-layer design, link quality/channel availability tradeoff and interference management are the key design principles for providing efficient spectrum utilization in CRNs. We start by describing the main CRN architectures and operating environment. Then, the spectrum sharing problem is stated. The various objectives used to formulate the spectrum sharing problem in CRNs are summarized. We then point out the several design challenges in designing efficient spectrum sharing and access mechanisms. The tradeoffs in selecting the operating channel(s) in CRNS are discussed. A number of spectrum sharing design categories are then surveyed. Various complementary approaches, new technologies and optimization methods that have great potential in facilitating the design of efficient CRN communication protocols are highlighted and discussed. Finally, concluding remarks are provided with several open research challenges.


\begin{figure}[h!]
\begin{center}
\epsfxsize=3in \epsfysize=4.5in \leavevmode
\epsfbox{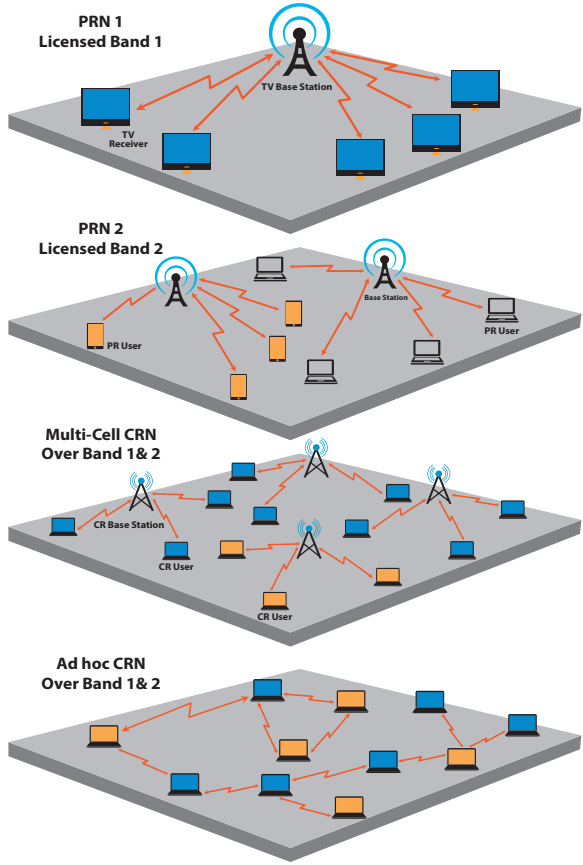} \epsfxsize=1.5in
\epsfysize=1.2in \caption{Generic architecture of a CRN environment.}
 \label{fig:CRNc}
\end{center} \end{figure}
\section{Network Architecture}
\subsection{CRN Model}
Typical CRN environment consists of a number of different types of PRNs and one or several CRNs. The PR and CR networks geographically co-exist
within the same area. In terms of network topology, two basic types of CRNs are proposed: centralized multi-cell CRNs and infrastructure-less ad hoc CRNs. Figure \ref{fig:CRNc} depicts a composition view of a CRN operating environment consisting of an ad hoc CRN and a multi-cell centralized CRN that coexist with two different types of PRNs. The different PRNs have license to transmit over orthogonal non-overlapping spectrum bands, each with a different licensed bandwidth. PR users of a given PRN operate over the same set of licensed channels. CR users can opportunistically utilize the entire PR licensed and unlicensed spectrum. For ad hoc multi-hop CRNs without centralized entity, it is necessary to provide distributed spectrum access protocols that allow each CR user to separately access and utilize the available spectrum. Furthermore, for centralized multi-cell CRNs, it is desirable to provide (1) centralized spectrum allocation protocols that allocate the available channels to the different CR cells, and (2) centralized channel assignment mechanisms that enable efficient spectrum reuse inside each cell.


\subsection{PR ON/OFF Channel Model}
\begin{figure}[htb]
\begin{center}
\epsfxsize=3.2in \epsfysize=2.0in \leavevmode
\epsfbox{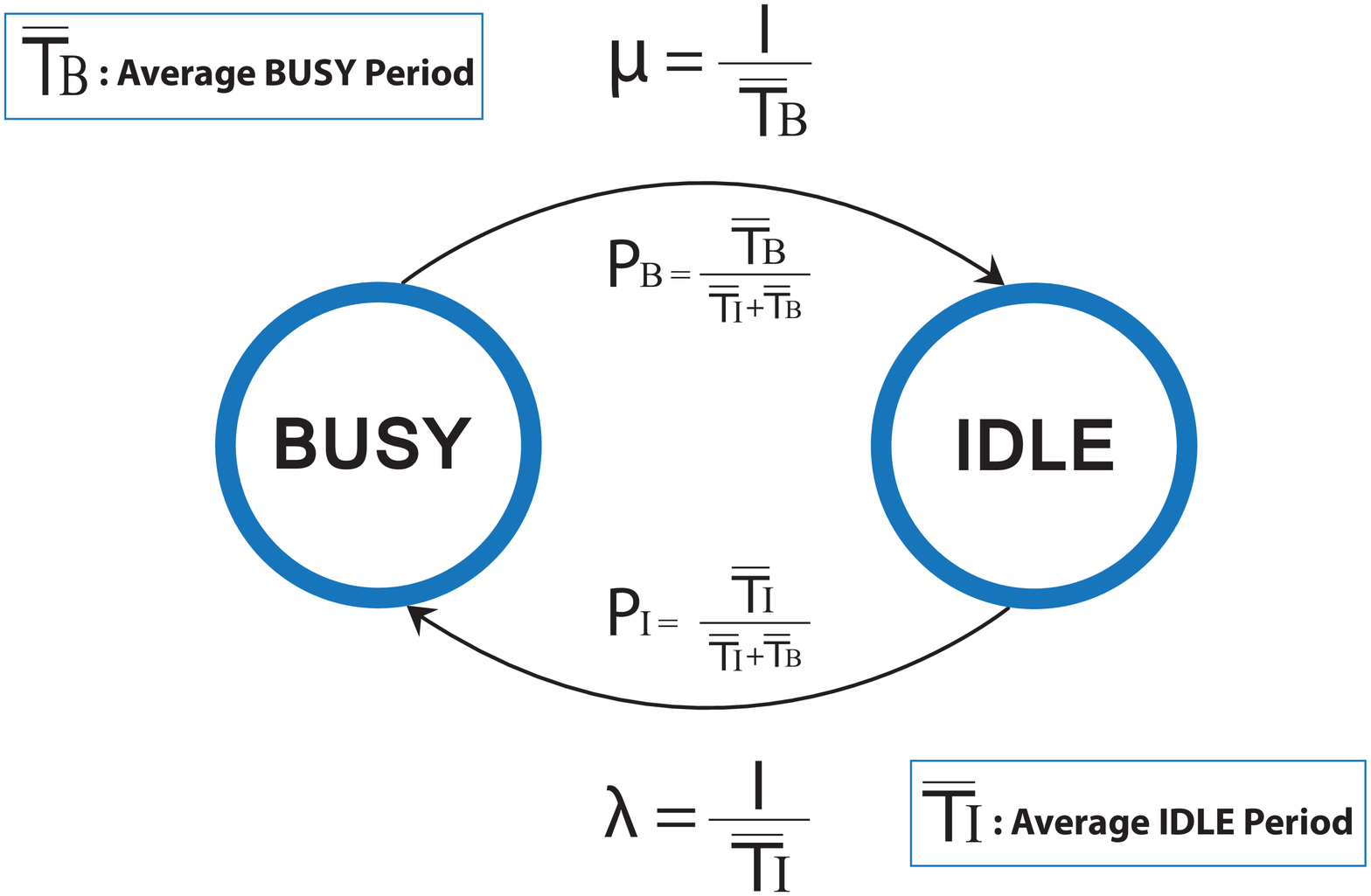} \epsfxsize=1.5in
\epsfysize=1.2in \caption{Two-state Markov channel availability model of a given PR channel.}
 \label{fig:ONOFF}
\end{center} \end{figure}

In general, the channel availability model of each PR channel in a given locality is described by a two-state ON/OFF Markov process. This model describes the evolution between idle (OFF) and busy (ON) states (i.e., the ON state of a PR channel indicates that the PR channel is busy, while the OFF state reveals that the PR channel is idle). The model is further described by the stochastic distribution of the busy and idle periods, which are generally distributed. The distributions of the idle and busy states depend on the PR activities.  We note here the ON and OFF periods of a given channel are independent random variables. For a given channel $i$, the average idle and busy periods are $\overline{T}_{I}$ and $\overline{T}_{B}$, respectively. Based on this model, the idle and busy probabilities of a PR channel $i$ are respectively given by $P_I^{(i)}=\frac{\overline{T}_{I}} {\overline{T}_{I}+\overline{T}_{B}}$ and $P_B^{(i)}=\frac{\overline{T}_{B}} {\overline{T}_{I}+\overline{T}_{B}}$. Figure \ref{fig:ONOFF} shows a transition diagram of a $2$-state busy/idle Markov model of a given PR channel. We note here that neighboring CR users typically have similar views to spectrum availabilities, while non-neighboring CR users have different channel availability conditions.


\section{Spectrum Sharing Problem Statement and Objectives}
Spectrum sharing problem (including spectrum management and decision) can be stated as follows: `` Given the output of spectrum sensing, the main goal is to determine which channel(s) to use, at what powers, and at what rates? such that a given performance metric (objective function) is optimized. This is often a joint optimization problem that is very difficult to solve (often it constitutes an NP-hard problem). Recently, several spectrum assignment strategies have been proposed for CRNs
\cite{[63Go],[72Go],[73Go],[11],[37],[75Go],[70Go],[3av],[74Go],[43FT],[71Go],[Haythem1],[79Go],[Haythem2],[Haythem3],[69SD],[82SD],[76Go],[77Go],[78Go],[80Go],[53EGo],[54EGo],[91EGo],[92EGo],[71F],[75F],[77F],[78F],[95Co],[96Co],[97Co]}. These strategies are designed to optimize a number of performance
metrics including:

\begin{itemize}
  \item Maximizing the CR throughput (individual users or network-level) based on Shannon capacity or a realistic staircase rate-SINR function (e.g., \cite{[63Go],[72Go],[73Go]}).
   \item Minimizing the number of assigned channels for each CR transmission (e.g., \cite{[11],[37]}).
 \item Maximizing the CR load balance over the different PR channels (e.g., \cite{[75Go]}).
  \item Minimizing the probability of PR disruption, i.e., minimizing the PR outage probability (e.g., \cite{[70Go],[3av]}).
    \item Minimizing the average holding time of selected channels (minimizing the PR disruption) (e.g., \cite{[74Go],[43FT]}).
    \item Minimizing the frequency of channel switching due to PR appearance by selecting the channel with maximum residual idle time, i.e., minimizing the CR disruption in terms of forced-termination rate (e.g., \cite{[71Go],[Haythem1],[79Go]}).
        \item Maximizing the CR probability of success (e.g., \cite{[Haythem2],[Haythem3]}).
  \item Minimizing the spectrum switching delay for CR users (e.g., \cite{[69SD],[82SD],[76Go]}).
  \item Minimizing the expected CR waiting time to access a PR channel (e.g., \cite{[77Go],[78Go]}).
  \item Minimizing CRN overhead and providing CR QoS (e.g., \cite{[80Go]}).
       \item Minimizing the overall energy consumption (e.g., \cite{[53EGo],[54EGo],[91EGo],[92EGo]}).
    \item Achieving fair spectrum allocation and throughput distribution in the CRN (e.g., \cite{[71F],[75F],[77F],[78F]}).
  \item Maintaining CRN connectivity with predefined QoS requirements (e.g., \cite{[95Co],[96Co],[97Co]}).
\end{itemize}

We note here that the spectrum sharing problem for any of the aforementioned objectives is, in general, NP-hard.
Therefore, several heuristics algorithms and approximations have been proposed to provide suboptimal solutions for the problem in polynomial-time.
These heuristics and approximations can be classified based on their adopted optimization method as: graph theory-based algorithms (e.g., \cite{[134GRAph]}), game theory-based algorithms (e.g., \cite{[82SD]}), genetic-based algorithms (e.g., \cite{[Genetic]}), linear programming relaxation-based algorithms (e.g., \cite{[66Go]}), fuzzy logic-based algorithms (e.g., \cite{[148FU]}), dynamic programming-based algorithms (e.g., \cite{[66DY]}), sequential-fixing-based algorithms (e.g., \cite{[37]}).

\section{Issues in Designing Spectrum Sharing Mechanisms}

\subsection{Interference Management and Co-existence Issue}
The coexistence problem is one of the most limiting factors in achieving efficient CR communications. In a CRN environment, there are three kinds of harmful interference that should be considered: PR-to-CR/CR-to-PR interference (the so-called PR coexistence) and CR-to-CR interference (the so-called self-coexistence). While several mechanisms have been proposed to effectively deal with the PR-to-CR interference problem based on cooperative (e.g.,\cite{[Sensing2],[Sensing3],[Sensing4],[Sensing5]}) or noncooperative (e.g.,\cite{[Sensing1],[Sensing11],[Sensing12]}) spectrum sensing, the CR-to-CR and CR-to-PR interference problems are still challenging issues.
\subsubsection{Self-coexistence Management}
To address the CR-to-CR interference problem in ad hoc CRNs, several channel allocation and self-coexistence management mechanisms have been proposed based on either (1) exclusive channel assignment or (2) joint channel assignment and power control. On the contrary, the CR-to-CR interference problem has been addressed in multi-cell centralized CRNs based on either fixed channel allocation~\cite{[8WRAN],[9WRAN],[10WRAN],[11WRAN],[7WRAN]} or adaptive traffic-aware/spectrum-aware channel allocation~\cite{[9WRANH],[10WRANH],[11WRANH]}.

\subsubsection{Providing Performance Guarantees to PR users}
It has been shown that the CR-to-PR interference is the most crucial interference in CRN environment, because it has a direct effect on the performance of PRNs. Hence, the transmission power of CR users over the PR channels should be adaptively computed such that the performance of the PRNs is protected. Based on the outcomes of spectrum sensing, two different power control strategies can be identified: binary and multi-level transmission power strategies. According to the binary-level strategy (the most widely used power control strategy in CRNs), CR users can only transmit over idle channels with no PR activities. Specifically, for a given PR channel, a CR transmits $0$ power if the channel is busy, and uses the maximum possible power if the PR channel is idle. While this strategy ensures collision-free spectrum sharing between the CR and PR users, it requires perfect spectrum sensing. Worse yet, the binary-level strategy can lead to non-optimal spectrum utilization. On the other hand, using a multi-level adaptive frequency-dependent transmission power strategy allows the CR and PR users to simultaneously share the available spectrum in the same locality, which can significantly improve spectrum utilization. By allowing CR users to utilize both idle and partially-occupied PR channels, much better spectrum utilization can be achieved. The multi-level power strategy can also be made time-dependent to capture the dynamics of PR activities. Under this strategy, controlling the CR-to-PR interference is nontrivial. In addition, computing the appropriate and adequate multi-level power strategy is still a challenging issue, which has been studied under some simplified assumptions. Specifically, the authors in \cite{[11]} proposed an adaptive multi-level frequency- and locality-dependent CR transmission power strategy that provides a soft guarantee on PRNs' performance. This adaptive strategy is dynamically determined according to the PR traffic activities and interference margins.


\subsection{Distributed Coordination Issue}
\label{sec:related}
In this section, we review several well-known distributed coordination mechanisms designed for CRNs. We note that control channel designs for CRNs can be loosely classified into six different categories~\cite{[12],[13]}:
\begin{itemize}
  \item Dedicated pre-specified out-of-band common control channel (CCC) design \cite{[3av],[3],[8],[8a],[9]}.
  \item Non-dedicated pre-specified in-band CCC design \cite{[16a],[Haythem3],[5av],[15]}.
  \item  Hybrid out-of-band and in-band CCC design \cite{[HYBRID]}.
  \item Hopping-based control channel design \cite{[79Gox],[18],[19],[22],[23],[24],[25]}.
  \item Spread-spectrum-based control channel design \cite{[DSSS],[DSSS1]}.
  \item Cluster-based local CCC design~\cite{[27],[27a],[28],[38],[29]}.
\end{itemize}
Despite the fact that using a dedicated out-of-band CCC is straightforward, it contradicts the opportunistic behavior of CRNs, may result in a single-point-of-failure (SPOF) and performance bottleneck due to CCC saturation under high CR traffic loads. Similarly, using a pre-specified non-dedicated in-band CCC is not a practical solution due to spectrum heterogeneity and, if exists, such solution can result in a SPOF, become a performance bottleneck, and introduce security challenges. Another approach that can effectively deal with the CCC saturation issue (bottleneck problem) is to use a hybrid out-of-band and in-band CCC (simultaneous control communications over in-band PR channels and dedicated out-of-band CCCs). This approach exploits the strengths of out-of-band and in-band signaling and, hence, can significantly enhance the performance of multi-hop CRNs. Using a hopping-based control channel can address the SPOF, bottleneck and security issues. However, in such type of solutions, the response to PR appearance is challenging as CR users cannot use a PR channel once reclaimed by PR users. In addition, this type of solution is generally spectrum unaware. Another key design issue in such solutions is the communication delay that heavily depends on the time to rendezvous. Using cluster-based coordination solutions, where neighboring CR users are dynamically grouped into clusters and establish local CCCs, can provide reliable distributed coordination in CRNs~\cite{[13],[12]}. However, adopting this type of solutions in a multi-hop CRN is limited by several challenges, such as providing reliable inter-cluster communication (i.e., different cluster may consider different CCCs), maintaining connectivity, broadcasting control information, identifying the best/optimal cluster size, and maintaining time-synchronization \cite{[13]}. Finally, using spread-spectrum-based distributed coordination is a promising solution to most of the aforementioned design challenges, but the practicality and design issues of such solution need to be further investigated. According to this solution, the control information is spread over a huge PR bandwidth with a very low transmission power level (below the noise level). Consequently, with a proper design, an efficient CCC design can be implemented using spread spectrum with minor effect on PRNs' performance. In conclusion, various distributed coordination mechanisms have been developed to provide reliable communications for CRNs, none of which are totally satisfactory. Hence, designing efficient distributed coordination schemes in CRNs should be based on novel coordination mechanisms along with effective transmission technologies that enable effective, robust and efficient control message exchanges.

\section{Tradeoffs in Selecting the Operating Channel}

The spectrum (channel) assignment problem in CRNs has been extensively studied in the literature. Existing channel assignment/selection solutions can loosely be classified into three categories: best link-quality schemes, larger availability-period schemes, and joint link-quality and channel-availability-aware schemes. It has been shown (e.g.,\cite{[3av],[2av],[Haythem1]}) that using the best link-quality schemes in CRNs, where the idle channel(s) with the highest transmission rate(s) are selected, can only provide good performance under relatively static PR activities with average PR channel idle durations that are much larger than the needed transmission times for CR users \cite{[3av],[5av],[Haythem1],[Haythem2],[Haythem3]}. Under highly dynamic PR activities, this class of schemes can result in increasing the CR termination rate, leading to a reduction in CRN performance as a CR user may transmit over a good-quality PR channel with relatively short availability time (short channel-idle period). On the other hand, employing the larger availability-period schemes in CRNs (e.g., \cite{[6av]}) can result in increasing the CR forced-termination rate as an idle PR channel of very poor link-quality (low transmission rate) may be chosen, resulting in a significant reduction in CRN performance. We note here that the interaction between the CRN and PRNs is fundamental for conducting channel assignment in CRNs.

The above discussion presents sufficient motivation to jointly consider the link-quality and average idle durations of PR channels when assigning operating channels to CR users. However, several open questions in this domain still need to be addressed; possibly the most challenging one is how to jointly consider the link-quality and average idle durations into one metric to perform channel assignment. Other important questions are: How can a CR user estimate the distribution of the idle periods of the different PR channels? What are the implications of the interaction between the CRN and the PRNs? How can a CR user determine the link-quality conditions over the various (large number) PR channels? Some of these questions have been addressed in \cite{[Haythem1],[Haythem2],[Haythem3]} by introducing the CR packet success probability metric. This metric is derived based on stochastic models of the time-varying PR behaviors. The probability of success over a given channel is a function of both the link-quality condition and the average-idle period of that PR channel. It has been proven that it is necessary to jointly consider the link-quality conditions and availability times of available PR channels to improve the overall network performance~\cite{[Haythem1]}.
%
%
%

\section{State-of-the-Art Spectrum Sharing Protocols in CRNs}
There are several attempts have been made to design spectrum sharing protocols with the objective of improving the overall spectrum utilization while protecting the performance of licensed PRNs. Existing spectrum sharing/access protocols and schemes for CRNs can loosely be categorized into four main classes based on: the number of radio transceivers per CR user (single-transceiver, dual-transceiver, or multiple transceiver), their reaction to PR behavior (reactive, proactive, or interference-based threshold), their spectrum allocation behavior (exclusive or non-exclusive spectrum occupancy model), and the guardband considerations (guardband-aware or guardband-unaware).

\subsection{Number of Radio Transceivers and Assigned Channels}
Spectrum sharing protocols and schemes for CRNs can also be categorized based on the number of radio transceivers per a CR user (i.e., single transceiver \cite{[116],[120],[4SR],[129],[127],[131],[128]}, dual transceivers~\cite{[124],[147]}, and multiple transceivers~\cite{[3av],[5av],[Haythem2],[2MC]}). Using multiple (or dual) transceivers greatly simplifies the task of spectrum access design and significantly improve system performance. This is because a CR user can simultaneously utilize multiple channels (the potential benefits of utilizing multi-channel parallel transmission in CRNs were demonstrated in \cite{[37],[Taos]}). In addition, the spectrum access issues such as hidden/exposed terminals, transmitter deafness and connectivity can be easily overcome as one of the transceivers can be switched to the assigned control channel (i.e., CR users can always receive control packet over the CCC even when they are operating over the data channels). However, the achieved performance gain of using multiple transceivers (multi-channel parallel transmission) comes at the expense of extra hardware. Worse yet, the optimal joint channel assignment and power control problem in multi-transceiver CRNs is, in general, NP-Hard. On the other hand, it has been shown that the design of efficient channel assignment schemes
for single-transceiver single-channel low-cost CRNs is simpler than that of the multi-transceiver counterpart \cite{[5av]}. While single-transceiver designs can greatly simplify the task of finding the optimal channel assignment, the aforementioned channel access issues are not trivial and the performance is limited to the capacity of the selected channel.



\subsection{Reaction to PR Appearance}
Spectrum sharing schemes in the CRNs can also be classified based on their reaction to the appearance of PR users into three main groups: (1) proactive (e.g.,  \cite{[83Pro],[84Pro],[86Pro]}), (2) reactive (e.g., \cite{[91Reactive],[87Reactive],[88Reactive]}), and (3) interference threshold-based (e.g.,~\cite{[Taos],[37],[11],[68Inter]}). In reactive schemes, the active CR users switch channels after the PR appearance. On the other hand, in proactive schemes, the CR users predict the PR appearance and switch channels accordingly. The threshold-based schemes allow the CR users to share the spectrum (both idle and partially-occupied PR channels) with PR users as long as the interference introduced at the PR users is within acceptable values. Existing threshold-based schemes attempt at reducing the impacts of the un-controllable frequency-dependent PR-to-CR interference on CRN performance through proper power control based on either (1) the instantaneous sensed interference~\cite{[37]}, (2) the average measured PR interference \cite{[Taos]}, or (3) using stochastic PR interference models \cite{[11]}.

\subsection{Spectrum Sharing Model}
The spectrum sharing model represents the type of interference model used to solve the channel and power assignment problem. There are two different spectrum sharing models: protocol (interference avoidance) and physical (interference) models~\cite{[13]}. The former employs an exclusive channel occupancy strategy, which eliminates the CR-to-CR interference and simplifies the management of the CR-to-PR interference \cite{[37],[3av]}. However, it does not support concurrent CR transmissions over the same channel, which may reduce the spectrum efficiency. On the other hand, the overlay physical model allows for multiple concurrent interference-limited CR transmissions to simultaneously proceed over the same channel in the same locality, which improves spectrum efficiency \cite{[FAN]}. However, the power control issue (CR-to-CR and CR-to-PR interference management) under this model is not trivial. Worse yet, using this model requires a distributed iterative power adjustment for individual CR users, which was shown that it results in slow protocol convergence~\cite{[FAN]}.

\subsection{Guard-band Considerations}
Most of existing spectrum sharing protocols for CRNs were designed assuming orthogonal channels, where the adjacent channel interference (ACI) is ignored (e.g., \cite{[16a],[Haythem3],[3av],[5av],[116],[9],[35]}). However, this requires using ideal sharp transmit and receive filters, which is practically not feasible. In practice, frequency separation (guard bands) between adjacent channels is needed to mitigate the effects of ACI and protect the performance of ongoing PR and CR users operating over adjacent channels. It has been shown that introducing guard bands can significantly impact the spectrum efficiency, and hence it is very important to account for the
guard-band constraints when designing spectrum sharing protocols for CRNs.

Few number of CRN spectrum access and sharing protocols have been designed
while accounting for the guard band issue \cite{[37],[8gb],[9gb]}.
Guard band-aware strategies enable effective and safe spectrum sharing, have a great potential to enhance the spectral efficiency, and protect the receptions of the ongoing CR and PR transmissions over adjacent channels. The need for guard band-aware spectrum sharing mechanisms and protocols was discussed in \cite{[37]}. Specifically, the authors, in\cite{[37]}, have investigated the ACI problem and proposed guard-band-aware spectrum access/sharing protocols for CRNs. The main objective of their proposed mechanism is to minimize the total number of reserved guard-band channels such that the overall spectrum utilization is maximized.  In \cite{[8gb]}, the authors showed that selecting the operating channels on per block (group of adjacent channels) basis instead of per channel basis (unlike the work in \cite{[37]}) provides better spectrum efficiency. The work in \cite{[8gb]} attempts at selecting channels such that at most one guard band is introduced for each new CR transmission. In \cite{[9gb]}, the authors proposed two guard-band spectrum sharing mechanisms for CRNs. The first mechanism is a static single-stage channel assignment that is suitable for distributed multi-hop CRNs. The second one is an adaptive two-stage channel assignment that is suitable for centralized CRNs. The main objective of the proposed mechanisms is to maximize spectrum efficiency while providing soft guarantees on CR performance in terms of a pre-specified rate demand.

%
%

%

\section{Complementary Techniques and Optimization Methods}
In this section, we discuss and explain several methods and optimizations that interact with spectrum sharing protocols to further improve spectrum utilization in CRNs.

\subsection{Resource Virtualization in CRNs}
The resource virtualization concept has been extensively discussed in the literature, which refers to the process of creating
a number of logical resources based on the set of all available physical resources. This concept allows the users
to utilize the logical resources in the same way they are using the physical resources. This leads to a better utilization of the physical resources as virtualization allows more users to share the available physical resources. In addition, virtualization introduces an additional layer of security as user's application cannot directly control the physical resources. The concept of virtualization was originally used in computer systems to better utilize the available physical resources (e.g., processors, memory, storage units, and network interfaces). These resources are virtualized into separate sets of logical resources, each set of these virtual resources can be assigned to different users. Using system virtualization can achieve: (1) users' isolation, (2) customized services, and (3) improved resource efficiency. Virtualization was also been introduced in wired
networks by introducing the framework of virtual private networks (VPNs).

Recently, several attempts have been made to implement the virtualization concept in wireless CRNs. We note here that employing virtualization in CRNs is daunted by several challenges including: spectrum sharing, limited infrastructure, different geographical regions, self co-existence, PR co-existence, dynamic spectrum availability, spectrum heterogeneity, and users' mobility~\cite{[7v]}. In \cite{[SDS1]}, a single cell CRN virtualization framework was introduced. According to this framework, a network with one BS and $M$ physical radio nodes (PNs) with varying sets of resources are considered. The resources include the number of radio interfaces at each PN, the set of orthogonal idle channels at each PN, and the employed coding schemes. Each PN hosts a set of virtual nodes (VNs). The VNs located in the different PNs can communicate with each other. To facilitate such communications, VNs request resources from their hosting PNs. Simulation results have demonstrated the effectiveness of using network visrtualization in improving network performance. In \cite{[SDS2]}, the authors have proposed a virtualization framework for multi-channel multi-cell CRNs. In this work, a virtualization based semi-decentralized resource
allocation mechanism for CRNs using the concept of multilayer hypervisors was proposed. The main objective of this work is to reduce the overall CR control
overhead by minimizing the CR users' reliance on the base-station in assigning spectrum resources. Simulation results have indicated significant improvement in CRN performance (in terms of control overhead, spectrum utilization and blocking rate) is achieved by the virtualized framework compared to non-virtualized resource allocation schemes.

\subsection{Full Duplex Communications}

The problem of computing the optimal spectrum access strategy for CR users has been well investigated in \cite{[14xx],[17xx],[15xx]}, but for CR users that are equipped with half-duplex (HD) transceivers. It has been shown that using HD transceivers can significantly reduce the achieved network performance~\cite{[10xx]}. Motivated by the recent advances in full-duplex (FD) communications and self-interference suppression (SIS) techniques, several attempts have been made to exploit
the FD capabilities and SIS techniques in designing communication protocols for CRNs ~\cite{[10xx],[11xx],[12xx]}.
The main objective of these protocols is to improve the overall spectrum efficiency by allowing simultaneous transmission and reception (over the same channel or over different channels) at each CR user. These protocols, however, require additional hardware
support (i.e., duplexers). The practical aspects of using FD radios in CRNs need to be further investigated. The design of effective channel/power/rate assignment schemes for FD-based CRNs is still an open problem.

\subsection{Beamforming Techniques}
Beamforming techniques are another optimization that can enable efficient spectrum sharing~\cite{[Beamforming2],[Beamforming3],[Beamforming4],[Beamforming5]}.
According to beamforming, the transmit and receive beamforming coefficients are adaptively computed by each CR user such that the achieved CR throughput is maximized while minimizing the introduced interference at the CR and PR users. Furthermore, the performance gain achieved by using beamforming in CRNs can be significantly improved by allowing for adaptive adjustment of the allocated powers to the transmit beamforming weights \cite{[Beamforming5]}. The operation details of such an approach need to be further explored.


\subsection{Software Defined Radios and Variable Spectrum-width}
The use of variable channel widths through channel aggregation and bonding is another promising approach in improving spectral efficiency. However, this approach has not given enough attention. Based on its demonstrated excellent performance (compared to using fixed-bandwidth channels), variable channel widths
has been chosen as an effective spectrum allocation mechanism in cellular mobile communication systems, including the recently deployed 4G wireless systems. Thus, it is very important to use variable-bandwidth channels in CRNs. More specifically,
in CRNs, assigning variable bandwidth to different CR users can be achieved through channel bonding and aggregation. This has a great potential in improving spectrum efficiency. The use of variable bandwidth transmission in CRNs is not straightforward due to the dynamic time-variant behavior of PR activities and the hardware nature of most of existing CR devices \cite{[13]}, which make it very hard to control the channel bandwidth~\cite{[37]}.

So far, most of CR systems have been designed with the assumption that each CR user is equipped with single or several radio transceivers. Using hardware radio transceivers can limit the number of possibly assigned channels to CR users and cannot fully support variable-width channel assignment. One possible approach to enable variable-width spectrum assignment and increase network throughput is to employ software defined radios (SDRs). The use of the SDRs enables the CR users to bond and/or aggregate any number of channels, thus enabling variable spectrum-width CR transmissions. Thus, SDRs support more efficient spectrum utilization, which significantly improves the overall CRN performance and provides QoS guarantees to CR users.


\subsection{Cross-layer Design Principle}

Cross-layer design is essential for efficient operation of CRNs. Spectrum sharing protocols for CRNs should select the next-hop and the operating PR frequency channel(s) using a cross-layer design that incorporates the network, MAC, and physical layers. A cross-layer routing metric called the maximum probability of success (MPoS) was proposed in \cite{[Haythem3]}. The MPoS incorporates the link quality conditions and the average availability periods of PR users to improve the CRN performance in terms of the network throughput. The metric assigns operating channels to the candidate routes so that a route with the maximum probability of success and minimum CR forced termination-rate is selected. The main drawback of the MPoS approach is its requirement of known PR channel availability distributions (the probability density function of idle periods of the PR channels).

\subsection{Discontinuous-OFDM Technology}
Based on the spectrum availability conditions and to enable efficient CRN operation, a CR user may need to utilize multiple adjacent (contiguous) idle PR channels (the so-called spectrum bonding) or non-adjacent (non-contiguous) idle PR channels (the so-called spectrum aggregation). Spectrum bonding and aggregation can be realized using either the traditional frequency division multiplexing (FDM) or the discontinuous-orthogonal frequency division multiplexing  (D-OFDM) technology \cite{[37],[37v],[36ofdn]}. The former technology requires several half-duplex transceivers and tunable filters at each CR user, where each assigned channel will use one of the available transceivers. While this approach is simple, it requires a large number of transceivers and does not  provide the enough flexibility needed to implement channel aggregation and bonding at a large-scale. The D-OFDM is a novel wireless radio technology that allows a CR transmission to simultaneously take place over several (adjacent or non-adjacent) channels using one half-duplex OFDM transceiver. According to D-OFDM, each channel includes a distinct equal-size group of adjacent OFDM sub-carriers. According to D-OFDM, spectrum bonding and aggregation with any number of channels can be realized through power control, in which the sub-carriers of a non-assigned channel will be assigned $0$ power and all the sub-carries of a selected channel will be assigned controlled levels of powers. We note here that the problem of assigning different powers to different OFDM symbols within the same channel is still an open issue.

\subsection{Spectrum Sharing for MIMO-Based Ad Hoc CRNs}
Multiple Input Multiple Output (MIMO) is considered as a key technology to increase the achieved wireless performance. Specifically, MIMO can be used to improve spectrum efficiency, throughput performance, wireless capacity, network connectivity, and energy efficiency. The majority of previously proposed works on MIMO-based CRNs (e.g., \cite{[15MIMO],[16MIMO],[10MIMO]}) have focused on the physical layer and addressed a few of the challenging issues at the upper layers, but certainly more effort is still required to investigate the achieved capacity of MIMO-based CRNs, the design of optimal channel/power/rate assignment for such CRNs, the interoperability with the non-MIMO CRNs, and many other challenging issues.

\subsection{Cooperative CR Communication (Virtual MIMO)}
One of the main challenges in the design of CRNs communication protocols is the time-varying
nature of the wireless channels due to the PR activities and the multi-path fading. Cooperative communication is a promising approach that can deal with the time-varying nature of the wireless channels, and hence improve the CRN performance. Cooperative communication can create a virtual MIMO system by allowing CR users to assist each other in data delivery (by relaying data packets to the receiver). Hence, the received data packets at the CR destination traverse
several independent paths achieving diversity gains. Cooperative communication can also extend the coverage area. The benefits of employing cooperative communication, however, are achieved at the cost of an increase in power consumption, an increase in computation resources and an increase in system complexity. It has been shown that cooperation may potentially lead to significant long-term resource savings for the whole CRN. An important challenge in this domain is how to design effective cooperative MAC protocols that combine the cooperative communication with CR multiple-channel capability such that the overall network performance is improved. The CR relay selection is another challenging problem that needs to be further investigated. Therefore, new cooperative CRN MAC protocols and relay selection strategies are needed to effectively utilize the available resources and maximize network performance.

\subsection{Network Coding}
Network coding in CRNs is another interesting approach that has not yet explored in CRNs. Based on its verified excellent performance in wireless networks \cite{[NETCOD]}, it is natural to consider it in the design of cooperative-based CRNs. The packet relaying strategies in cooperative communication are generally implemented on a per packet basis, where a store-and-forward (SF) technique is used (the received packets at the CR relays are received, stored and retransmitted towards the receiver). While this type of relaying mechanisms is simple, it has been shown that it provides a sub-optimal performance in terms of the overall achieved CRN throughput (especially, in multi-cast scenarios). Instead of using SF, network coding can be used to maximize the CRN performance. With network coding, the intermediate relay CR users can combine the incoming packets using mathematical operations (additions and subtractions over finite fields) to generate the output packets.

One drawback in using network coding is that the computational complexity increases as the finite field size increases. The higher the field size, the better is the network performance. However, the tradeoff should be further investigated and more efforts are required to identify and study the benefits and drawbacks of increasing the field-size in CRNs. In addition, the performance achieved through network coding can be
further enhanced in CRNs by dynamically adapting the total number of coded packets that need to be sent by the source CR user. Such adaptation adjustment is yet to be explored, which should be based on the PR activities, link loss rates, link correlations, and nodes' reachability.

%
%


\section{Summary and Open Research Problems}
CR technology has a great potential to enhance the overall spectrum efficiency. In this paper, we first highlighted the main existing CRN architectures. Then, we described the unique characteristics of their operating RF environment that need to be accounted for in designing efficient communication protocols and spectrum assignment mechanisms for these networks. We then surveyed several spectrum sharing approaches for CRNs. We showed that these approaches differ in their design objectives. Ideally, one would like to design a spectrum sharing solution that maximizes spectrum efficiency while causing no harmful interference to PR users. We showed that interference management (including self-coexistence and the PR coexistence) and distributed coordination are the main crucial issues in designing efficient spectrum sharing mechanisms. The key idea in the design of effective spectrum sharing and assignment protocols for CRNs is to jointly consider the PR activities and CR link-quality conditions.

The reaction to PR appearance is another important issue in designing spectrum sharing schemes for CRNs. Currently, most of spectrum sharing schemes are either reactive or proactive schemes. Interference threshold-based schemes are very promising, where more research should be conducted to explore their advantages and investigate their complexities. Another crucial and challenging problem is the incorporation of the guard-band constraints in the design of spectrum sharing schemes for CRNs. A huge amount of interference is leaked into the adjacent channels when guard bands are not used. This can significantly reduce spectrum efficiency and cause harmful interference to PR users. The effect of introducing guard-bands on the spectrum sharing design has not been well explored.

Many interesting open design issues still to be addressed. Variable-width spectrum sharing approach is very promising, but their design assumptions and feasibility should be carefully investigated. Resource virtualization is another important concept that can significantly improve the overall spectrum utilization. Beamforming and MIMO technology have recently been proposed as a means of maximizing spectrum efficiency. The use of beamforming in CRNs with MIMO capability can achieve significant improvement in spectrum efficiency. However, the spectrum sharing problem becomes more challenging due to the resurfacing of several design issues such as the determination of the beamforming weights, the joint channel assignment and power control, etc., which need to be further addressed. Research should focus also on the cooperative CR communication and cross-layer concepts. Using FD radios versus using HD radios is another interesting issue. Moreover, utilizing network coding is very promising in improving the CRN's performance. Finally, we showed that channel bonding and aggregation can be realized through the use of D-OFDM technology. This technology allows CR user to simultaneous transmit or receive over multiple channels using a single radio transceiver.


%
%
%
%
%
%

\end{document}